\documentclass[10pt,showpacs,amsmath,amssymb]{article}
\usepackage{amsfonts}
\usepackage{mathdots}
\newcommand{\field}[1]{\mathbb{#1}}

\begin{document}

\title{Canonical surfaces associated with projectors in grassmannian sigma models}

\author{V. Hussin${}^{1,2,4}$, {I}.\ Yurdu\c{s}en${}^{1,5}$ and W. J. Zakrzewski${}^{3,6}$}

\footnotetext[1]{Centre de Recherches Math\'ematiques,
Universit\'e de Montr\'eal, C.P. 6128, Succ.~Centre-ville,
Montr\'eal (Qu\'ebec) H3C 3J7, Canada.}
\footnotetext[2]{D\'epartement de Math\'ematiques et de
Statistique, Universit\'e de Montr\'eal, C.P. 6128,
Succ.~Centre-ville, Montr\'eal (Qu\'ebec) H3C 3J7, Canada.}
\footnotetext[3]{Department of Mathematical Sciences, University of Durham, Durham DH1 3LE, United Kingdom.}
\footnotetext[4]{email:hussin@dms.umontreal.ca}
\footnotetext[5]{email:yurdusen@crm.umontreal.ca}
\footnotetext[6]{email:w.j.zakrzewski@durham.ac.uk}

\date{\today}

\maketitle

\begin{abstract}
We discuss the construction of higher-dimensional surfaces based on the harmonic maps of $S^2$ into $CP^{N-1}$ 
and other grassmannians. We show that there are two ways of implementing this procedure - both based 
on the use of the relevant projectors. We study various properties of such projectors and show that the Gaussian 
curvature of these surfaces, in general, is not constant. We look in detail at the surfaces corresponding to the Veronese
sequence of such maps and show that for all of them this curvature is constant but its value depends on which mapping
is used in the construction of the surface.

\end{abstract}



\section{Introduction \label{intro}}
The idea of using Weierstrass construction to generate surfaces in various multidimensional spaces 
 was first presented by Konopelchenko et al${}^{1,2}$ and then generalised
in several papers${}^{3-8}$.

Recently${}^{5}$, this procedure involved the use of $CP^{N-1}$ harmonic maps to construct such surfaces.
When it was generalised to the supersymmetric case${}^{6}$ it was realised that 
the key step in the procedure involved a set of projectors constructed out of these maps.
This was expanded further in Ref 7.
In this approach, the coordinates of the surface are given by line integrals of various derivatives 
of the corresponding projectors. To make these quantities independent of the contours of these
integrals one has to require that these projectors not only correspond to harmonic maps ({\it ie} satisfy
their equations) but also satisfy some further equations (see a further discussion in Section 3).

However, one can bypass the construction of line integrals ({\it ie} relate the coordinates of the surface directly to the components of the projectors) and so broaden somewhat the class of surfaces that can be constructed.
In either approach the properties of the surface depend on the structure of the projectors and so in this paper we have decided to study the relevant projectors in a more systematic way. In either case the projectors are required to satisfy the equations of the harmonic maps (so that the induced metric on the surfaces is conformal).

 Hence, after a short section (Section 2) in which we describe the construction  and various properties of harmonic maps 
(see {\it eg} Ref 9) we present in Section 3 a detailed discussion of our derivation of the coordinates of these
surfaces from the corresponding projectors. As these surfaces naturally live in $R^{N^2-1}$ 
we discuss in Section 4 various constraints that these coordinates have to satisfy. The fact that the surfaces 
live in $R^{N^2-1}$ is associated with
the trace of the projectors being an integer which allows us to eliminate one of the coordinates of the surface. This elimination makes the formulae look somewhat clumsy so in this paper, in all the formulae we give,
we ignore this constraint bearing in mind, however, that this constraint is always there. 
 In Section 5 we consider
the special case of projectors corresponding to the Veronese sequence.
In this case some coordinates naturally vanish - leading to the surfaces which live in
lower dimensional spaces. We call such projectors "reduced" and we discuss their
form in some detail. Of course, even in these cases, there are further constraints
between the coordinates of the surface, as after all, all our surfaces are two dimensional.

In our discussion we consider projectors of any rank; hence when the rank is larger than 
one we are really dealing with grassmannian models. In Section 6, we look in detail at projectors which arise from $CP^{N-1}$ harmonic maps corresponding to $N$ odd and even and find many interesting connections between them. We then present a short subsection of the properties 
of the surfaces corresponding to these solutions. In general, the surfaces have non-constant
Gaussian curvature but for the Veronese sequence their curvature is constant ({\it ie} they 
are all spheres of different radii). 

We finish the paper with some conclusions.

\section{Classical $\field{C}P^{N-1}$ sigma model and its 
solutions
\label{briefintroclass}}
In order to keep the paper self-contained we briefly review the 
$\field{C}P^{N-1}$ sigma model and recall some of its basic properties, 
which will be used in the subsequent developments. For further details on 
this subject we refer the reader to Ref 9 and 10 and references 
therein. 

The $\field{C}P^{N-1}$ sigma model equations in Euclidean space are defined to be the 
stationary points of the energy functional${}^{9}$
\begin{eqnarray}
S=\frac{1}{4}\int_{\Omega}(D_{\mu} z)^{\dagger} (D_{\mu} z) 
d\xi d\bar{\xi}\,, \qquad z^{\dagger}\cdot z=1\,, \nonumber \\
\field{C} \ni \xi=\xi^1+i \xi^2 \rightarrow 
z=(z_0, z_1, \ldots, z_{N-1})^T \in \field{C}^N,
\label{action}
\end{eqnarray}
where $D_{\mu}$ denote covariant derivatives acting on 
$z:\Omega \rightarrow \field{C}P^{N-1}$, 
defined by 
\begin{equation}
D_{\mu}z =\partial_{\mu}z- (z^{\dagger}\cdot \partial_{\mu}z)z,
\qquad \partial_{\mu}=\partial_{\xi^{\mu}}\,,
\qquad \mu=1,2.
\label{covader}
\end{equation}
Here, $\Omega$ is an open, connected subset of  the complex plane $\field{C}$, 
$\xi$ and $\bar{\xi}$ are local coordinates in $\Omega$ and as usual 
the symbol $\dagger$ denotes Hermitian conjugation. The energy functional 
(\ref{action}) is invariant under global 
$U(N)$ gauge transformations $z\, \rightarrow z^{\prime} = U z$ where $U\in U(N)$, and also under the local $U(1)$ 
gauge transformations $z\, \rightarrow z^{\prime} = z e^{i\phi}$, 
where $\phi$ is a real-valued function depending on $\xi$ and $\bar{\xi}$. 

The Euler-Lagrange equations thus take the form 
\begin{equation}
D_{\mu} D_{\mu} z+ z (D_{\mu} z)^{\dagger} (D_{\mu} z)=0.
\end{equation}
In homogeneous coordinates  
\begin{equation}
z=\frac{f}{\vert f \vert}\,,\qquad f \in \field{C}^N,
\label{defz}
\end{equation}
they can be written in the form of the conservation law
\begin{equation}
\partial K - \bar{\partial} K^{\dagger}=0\,, 
\label{conservation}
\end{equation}
where $K$ is a $N \times N$ matrix of the form 
\begin{equation}
K=\frac{1}{\vert f \vert^2}
\left(\bar{\partial}f\otimes f^{\dagger}-f\otimes 
\bar{\partial}f^{\dagger}\right)
+\frac{f \otimes f^{\dagger}}{\vert f \vert^4}
\big(\bar{\partial}f^{\dagger}\cdot f - f^{\dagger}\cdot\bar{\partial}f
\big)\,.
\label{defmatK} 
\end{equation}
The symbols $\partial$ and $\bar{\partial}$ denote the standard derivatives 
with respect to $\xi$ and $\bar{\xi}$ respectively, {\it i.e.}
$\partial=\frac{1}{2}\left(\partial_{\xi^1}-i\partial_{\xi^2}\right)$.

All finite action solutions of the $\field{C}P^{N-1}$ sigma model (for 
$\Omega = S^2$) have been constructed by A. Din et al${}^{11}$ and R. Sasaki${}^{12}$. In that 
construction one gets three classes of solutions, namely (i) holomorphic 
(i.e. $\bar{\partial} f = 0$), (ii) antiholomorphic (i.e. ${\partial} f = 0$) 
and (iii) mixed. 

The mixed and antiholomorphic solutions can be determined from 
the holomorphic nonconstant functions by the following 
procedure${}^{9}$ . We construct, first, the operator $P_{+}$ which is defined by its action on vector-valued functions on $\field{C}^{N}$ as
\begin{eqnarray}
P_{+}: f \in \field{C}^N \rightarrow
P_{+}f=\partial f - \frac{f^{\dagger}\cdot \partial f }{\vert f \vert^2} f \,.
\label{operatorintermsoff}
\end{eqnarray}
Then starting from any nonconstant holomorphic function $f\in \field{C}^N$, the successive application, say $k$ times with $k \le N-1$, of the 
operator $P_{+}$ allows one to find $N-2$ mixed solutions $P_{+}^k f$, for $k=1, ..., N-2$,
and $P_{+}^{N-1} f$ gives rise to an antiholomorphic solution. Let us recall that $P_{+}^{N} f=0$. We generate this way
 an orthogonal basis of solutions in $\field{C}^N$, {\it i.e.}
\begin{equation}
f^{\dagger} \cdot P_{+}^j f = 0\,, \qquad (P_{+}^i f)^{\dagger} \cdot P_{+}^j f = 0\,, 
\quad
i \ne j\,,\quad i,j=1,..., N-1.
\label{orthogonalitypr}
\end{equation}

The gauge invariant projector 
formalism for the $\field{C}P^{N-1}$ sigma model is also well-known${}^{9}$.
 By defining a rank 1 orthogonal 
projector, for any $g\in \field{C}^{N}$, as
\begin{equation}
P = \frac{g \otimes g^{\dagger}}{\vert g \vert^2}\,,
\qquad
P^{\dagger}=P\,, \qquad P^2=P\,,
\label{projector}
\end{equation}
we see, in particular, that $P_{+} g=(I-P) \partial g$.
The energy functional (\ref{action}) can be expressed as 
\begin{equation}
S = \int_{\Omega} \rm{tr}(\partial P\, \bar{\partial} P)\, 
d\xi d\bar{\xi}\,, 
\label{actionintermsofproject}
\end{equation}
and the Euler-Lagrange equations become
\begin{equation}
[\partial \bar{\partial} P, P] = 0\,,
\label{EulLageqs}
\end{equation}
which could also be written as a conservation law 
(analogue of (\ref{conservation}))
\begin{equation}
\partial [\bar{\partial} P, P] + \bar{\partial} [\partial P, P]=0\,.
\label{conservation2}
\end{equation}

Of course, if now $g= f$ or $g= P_{+}^j f$, for $j=1, ..., N-1$ where $f$ is holomorphic, then the equations
(\ref{EulLageqs}) are automatically satisfied.

The projector formalism, can be used, among other things, for the construction 
of surfaces in $\field{R}^{N^2-1}$ obtained from the $\field{C}P^{N-1}$ sigma model. 
The equivalence of the Euler-Lagrange equations (\ref{conservation}) 
and of the set of Dirac-type equations given in Ref 1-3, 
whose solutions are used to construct surfaces, was given in Ref 4. Later on these ideas led to a general procedure 
for obtaining surfaces from these harmonic maps${}^{5}$ 
and it has been shown${}^{6, 13}$ that some of these surfaces are 
related to the projector $P_0$, constructed out of 
holomorphic solutions of the corresponding maps. Then in Ref 7 it was suggested to obtain new surfaces by constructing 
new projectors. For this purpose a sequence of projectors of the form  
\begin{equation}
P_k:= P(V_k) = \frac{V_k \otimes V_k^{\dagger}}
{V_k^{\dagger} \cdot V_k}\,,
\quad 
\mbox {where}
\,\,\,\,
V_k = {P}_+^k f\,,
\qquad
k =0,1,...,N-1\,,
\label{concproject}
\end{equation}
where $f$ is holomorphic, were constructed. 

These projectors satisfy $^{9}$, for $k=1,2,...,N-2$, 
\begin{equation}
\partial P_0 = \frac{P_{+} f \otimes f^{\dagger}}{ \vert  f \vert^2}, \quad \partial P_k = \frac{P_{+}^{k+1} f \otimes ({P}_+^k f)^{\dagger}}{ \vert {P}_+^k f \vert^2}- \frac{{P}_+^k f  \otimes  ({P}_+^{k-1}f)^{\dagger}}{ \vert {P}_+^{k-1}f\vert^2} \,,
\end{equation}
\begin{equation}
\hbox{tr}(\partial P_0 \bar\partial P_0) = \frac{\vert P_+  f\vert^2}{\vert  
 f\vert^2}, \quad \hbox{tr}(\partial P_k \bar\partial P_k) 
= \frac{\vert P_+^ {k+1}  f\vert^2}{\vert  P_+^k f\vert^2}  
+ \frac{\vert P_+^k f\vert^2}{\vert {P}_+^{k-1}f\vert^2} \,,
\label{trace1}
\end{equation}

and${}^{14}$
\begin{equation}
\label{ggy}
[\partial P_0, P_0]= \partial P_0, \quad  [\partial P_k, P_k]= \partial (P_k+ 2 \sum_{j=0}^{k-1}P_j).
\end{equation}

Moreover,  one can generate even more projectors 
by taking their sums\footnote{This way we are really using solutions of higher rank grassmannian models as 
will be discussed below.} and obtain new surfaces by this 
procedure as suggested in Ref 7. 
Thus, we could take the following linear combinations as our projector 
\begin{equation}
P = \sum_{i=0}^{N-2} \alpha_i P_i\,,
\label{lincomproject}
\end{equation}
where $\alpha_i$ are either $0$ or $1$. We take only a maximum of
$N-1$ of the $P_i$ due to the completeness relation 
\begin{equation}
\sum_{i=0}^{N-1} P_i = I\,.
\label{completenessrel}
\end{equation}

\section{Some aspects of the expressions for the surfaces obtained from 
the projector formalism 
\label{canonicalexp}}
\subsection{Preliminaries}

To generate surfaces in $R^{N^2-1}$ we can follow Ref 5-7 and define $X$, the coordinates of the surface,  by line integrals
 \begin{equation}
\label{newa} i \int_{\gamma} (\mathbb{K}^{\dagger} d\xi' +
\mathbb{K} d\bar{\xi'}) = X(\xi,\bar{\xi}).
\end{equation}
These expressions do not depend on the curve $\gamma$ but only on its endpoints (of which one is 
taken at $\infty$ and the other at ($\xi,\bar{\xi}$)) if 
\begin{equation}
\label{newb} 
d X = i (\mathbb{K}^{\dagger} \, d\xi + \mathbb{K}\, d\bar{\xi})= 
i(\mathbb{K}^{\dagger} +\mathbb{K})d\xi^1 -(\mathbb{K}^{\dagger} -\mathbb{K})d\xi^2 
\end{equation}
is closed $(d (d X)=0)$. This can be guaranteed if $\mathbb{K}$ and $\mathbb{K}^{\dagger}$ are chosen conveniently. In particular, in the previous work${}^{5}$
it was shown that given that the projectors which are the solutions of the equations of the $\field{C}P^{N-1}$ model satisfy (\ref{conservation2}) we can construct $\mathbb{K}$ out of various entries of such projectors.

However, the key role played by the projectors was fully understood when this procedure was generalised to the supersymmetric case${}^{9}$.
This was expended further in Ref 7.

The construction of surfaces in $\field{R}^{N^2-1}$ based on line integrals (\ref{newa})
involving the $\field{C}P^{N-1}$ sigma model 
has already  been  discussed in several papers${}^{6,7,8,13,14}$. 

Here we would like to make a few comments about this construction and then to suggest an alternative approach.
First if we take $P=P_0$  we observe that $d X$, constructed out of its entries, is closed and the expression for $X$ is independent of the contour of integration 
$\gamma$ in (\ref{newa}). If we consider other projectors $P_i$ for  $i\ne0$ we have to consider the sums given in (\ref{ggy}).

However,  we do not have to restrict our attention to the solutions of the $\field{C}P^{N-1}$ models. We can consider other 
grassmannians (described by sums of projectors). Thus we can take
\begin{equation}
\label{newc}
\mathbb{P}\,=\,\sum_{i=0}^{k} P_i\ ,\quad k=1,.., N-2.
\end{equation}
Such projectors describe the selfdual solutions of the grassmannian model  identified as the coset space $G(N,k+1)=U(N)/ ({U(N-k-1)\times U(k+1)})$, and so still satisfy
\begin{equation}
\label{newd}
[\partial \mathbb{P},\,\mathbb{P}]\,=\,\partial \mathbb{P}.
\end{equation}

However, all this is not necessary, if we are only interested in surfaces and their properties. We do not need to define
$X$ as line integrals of derivatives of projectors; we can take directly
\begin{equation}
\label{newe}
X\,=\,P_i
\end{equation}
or 
\begin{equation}
\label{newf}
X\,=\,\sum_i\,\alpha_i P_i,
\end{equation}
where $\alpha_i$ are arbitrary constants and $P_i$ do not even need to satisfy any particular equations ({\it eg} 
(\ref{EulLageqs})).

However, if we want to guarantee that the induced metric on the surface is conformal we require that the projectors $P_i$
solve (\ref{EulLageqs}). Then the induced metric $g$ will have only one nonvanishing component $g_{+-}$
and the Gausssian curvature will be  proportional to
\begin{equation}
\label{curve}
K\,=\,-\frac{4}{g_{+-}}\,\partial\bar\partial \ln(g_{+-}).
\end{equation}

So in this paper we restrict our attention to such cases.

\subsection{Some specific properties of projectors and of the corresponding surfaces}

To start, let us recall some general properties of orthogonal projectors which will be useful for the subsequent developments. A matrix  $P=(P_{ij})\in \field{C}^{N\times N}$ is called an orthogonal projector of order $N$ if we have $P^2=P$ and $P^{\dagger} = P$.  This means that $P$ is a Hermitian matrix with $P_{ii}\in \field{R}$ and ${\bar P}_{ij}=P_{ji}, \ i,j=1,...,N$ with $det \ P=0$. Any such projector thus takes the expression (\ref{lincomproject}), where the $P_i$ are orthogonal projectors of rank 1 which could be written as
\begin{equation}
P_i= {\hat u_i}\otimes {\hat u_i}^\dagger, \quad  {\hat u_i}^\dagger \cdot {\hat u_i}=1, \quad u_i \in \field{C}^{N}.
\label{rk1projector}
\end{equation}
The rank $r$ of a general orthogonal projector $P$, which can take the values $1,...,N-1$ and coincides with the trace, is thus equal to the number of $P_i$ appearing in a specific linear combination (\ref{lincomproject}).  

The constraint $P^2=P$  gives rise to a set of $N+ \frac{N(N-1)}{2}$ nonlinear equations between the entries of $P$:
\begin{eqnarray}
\sum_{j=1}^N |P_{ij}|^2 + P_{ii}(P_{ii}-1) = 0\,,
\qquad j \neq i\,, \quad i = 1, \ldots , N\,,
\label{genconst1}
\end{eqnarray}
\begin{eqnarray}
\sum_{m=1}^N P_{im} {\bar P}_{jm} +P_{ij}(P_{ii} + P_{jj}-1) = 0\,, 
\quad i < j,
\,\,
i \neq j \neq m,
\,\,\,
i,j = 1, \ldots , N.
\label{genconst2}
\end{eqnarray}

Starting now from  an orthogonal projector $P$ constructed as (\ref{lincomproject}) from solutions of the $\field{C}P^{N-1}$ sigma model, we give a procedure to get the coordinates of the radius vector in $\field{R}^{N^2-1}$. We know that a standard choice of a set of $N(N-1)$ real components of the radius vector $X$ is given by
\begin{equation}
(X_{ij})_+ = P_{ij} + \bar{P}_{ij},\quad
(X_{ij})_- = i(P_{ij} - \bar{P}_{ij}),\quad
i<j, \quad i,j=1,\ldots,  N ,
\label{xijplusminus}
\end{equation}
which will satisfy
\begin{equation}
\sum_{\begin{array}{c}
i,j=1 \\
i<j
\end{array}}^{N}\left( (X_{ij})_+^2+(X_{ij})_-^2\right) = 4\sum_{\begin{array}{c}
i,j=1 \\
i<j
\end{array}}^{N}|P_{ij}|^2.
\label{constraintforoffdiag}
\end{equation}
The remaining $(N-1)$ components of the radius vector $X$ are chosen as a linear combination of  the 
diagonal entries of $P$. They will be denoted as $X_1,\cdots X_{N-1}$. 
The relative freedom in the choice of these last components has lead  to different representations of the surface corresponding to the same solution of the $\field{C}P^{N-1}$ sigma model. Such a surface is characterised by a quadratic equation on the components of the radius vector $X$.

Let us exhibit here a canonical expression for this surface by imposing the following quadratic equation:
\begin{equation}
\sum_{i=1}^{N-1} X_i^2+ \sum_{\begin{array}{c}
i,j=1 \\
i<j
\end{array}}^{N}\left( (X_{ij})_+^2+(X_{ij})_-^2\right)  = C(N, r)\,,
\label{canforsur}
\end{equation}
where  $C(N ,r)$ is a constant depending on $N$ and on the rank $r$ of 
the projector $P$. The expression of the constant  $C(N ,r)$ as well as the components $X_1,\cdots, X_{N-1}$  thus have to be determined. 

Using (\ref{genconst1}), (\ref{constraintforoffdiag}) and $tr P=r$, (\ref{canforsur}) becomes
\begin{equation}
\sum_{i=1}^{N-1} X_i^2
= 2 (\sum_{i=1}^N P_{ii}^2 - r)+C(N,r) \,.
\label{inreachingcanexp1}
\end{equation}
The trace property  allows us to express $P_{NN}$ in terms of the other $P_{ii}$ and 
(\ref{inreachingcanexp1}) becomes
\begin{eqnarray}
\sum_{i=1}^{N-1}X_i^2 = 4\sum_{i=1}^{N-1}P_{ii}^2 + 
4\!\!\!\!\!\!\!\! \sum_{\begin{array}{c}
i,j=1 \\
i<j
\end{array}}^{N-1} \!\!\!\!\!\!P_{ii}P_{jj} - 
4r\sum_{i=1}^{N-1}P_{ii} + 2r (r-1) + C(N,r)\,.
\label{inreachingcanexp3}
\end{eqnarray}
We define a vector $\cal X$ constructed from the components  $X_1,\cdots, X_{N-1}$ and we express it as a linear combination of the independent diagonal entries of the matrix $P$. We thus get 
\begin{eqnarray}
{\cal X}:=\left(\begin{array}{c}
X_{1} \\
\vdots \\
X_{N-1}
\end{array}\right) = \ A {\cal P} +b= \ A \left(\begin{array}{c}
P_{11} \\
\vdots \\
P_{(N-1)(N-1)}
\end{array}\right) + b\,,
\label{definingx}
\end{eqnarray}
where
$A$ is a $(N-1) \times (N-1)$ matrix and $b$ is a $(N-1)$ vector still
to be determined. Indeed, using (\ref{inreachingcanexp3}) with (\ref{definingx}) we get
\begin{eqnarray}
{\cal X}^T {\cal X} = \mathcal{P}^T \left(\begin{array}{cccc}
4 & 2 & \cdots & 2 \\
2 & 4 & \, & \vdots \\
\vdots & \, & \ddots & 2 \\
2 & \cdots & 2 & 4
\end{array}\right)\mathcal{P} - 4r\mathcal{P}^T\left(\begin{array}{c}
1 \\
\vdots \\
1
\end{array}\right) \!\!+ 2r(r-1) + C(N,r),
\label{inreachingcanexp4}
\end{eqnarray}
which  implies that
\begin{eqnarray}
&&A^T A = \left(\begin{array}{cccc}
4 & 2 & \cdots & 2 \\
2 & 4 & \, & \vdots \\
\vdots & \, & \ddots & 2 \\
2 & \cdots & 2 & 4
\end{array}\right)\,,
\qquad
b=-2r(A^T)^{-1}\left(\begin{array}{c}
1 \\
\vdots \\
1
\end{array}\right)\,, \nonumber \\
\, \nonumber \\
&& \qquad \qquad \qquad \qquad \,\,\,\,
C(N,r) = b^T\,b - 2r(r-1)\,.
\label{finalexpforcanconst}
\end{eqnarray}
To compute $C(N,r)$, we need  the inverse of $A^T A$ which is easily found to be given by 
\begin{eqnarray}
(A^T A)^{-1} = \frac{1}{2N} \left(\begin{array}{cccc}
N-1 & -1 & \cdots & -1 \\
-1 & N-1 & \, & \vdots \\
\vdots & \, & \ddots & -1 \\
-1 & \cdots & -1 & N-1
\end{array}\right)\,,
\label{inverseofa}
\end{eqnarray}
and hence we see that
\begin{eqnarray}
C(N, r) = \frac{2r}{N}(N-r)\,.
\label{finalformforcostcan}
\end{eqnarray}

A canonical choice of the matrix $A$ would involve taking it  as a triangular matrix such that $X_{1} = P_{11}-P_{NN}$. In this case, we find
\begin{eqnarray}
\left\{
\begin{array}{cc}
A_{11} = 2\,, & \, \\
A_{1j} = 1\,, & \qquad j=2,3, \ldots,N-1, \\
A_{ij} = 0\,, & \quad {\rm if} \,\,\,\,i>j,
\end{array}\right. \nonumber \\
\left\{
\begin{array}{cc}
A_{ii} = \left(\frac{2(i+1)}{i}\right)^{{1}/{2}}\!\!\!\!\!\!, &
\!\!\!\!\!\!\!\! i = 2, \ldots, N-1\,, \\
A_{ij} = \left( \frac{2}{i(i+1)}\right)^{{1}/{2}}\!\!\!\!\!\!\!\!, & 
\quad
\begin{array}{c}
i = 2, \ldots ,N-1, \\
j = 3, \ldots , N-1,
\end{array}
\qquad i \neq j, \qquad i<j.
\end{array}\right.
\label{finalformformamatrix}
\end{eqnarray}
We also find that the components of the vector $b$ are 
\begin{equation}
b_i=-r \left(\frac{2}{i(i+1)}\right)^{{1}/{2}}, \  i = 1, \ldots, N-1.
\end{equation}
Finally, the canonical expression for the 
surface becomes 
\begin{eqnarray}
\sum_{i=1}^{N-1} X_i^2+ \sum_{\begin{array}{c}
i,j=1 \\
i<j
\end{array}}^{N}\left( (X_{ij})_+^2+(X_{ij})_-^2\right)   = \frac{2r}{N}(N-r)\,.
\label{forsrfaceeqfinalform}
\end{eqnarray}

The components $X_1,\cdots, X_{N-1}$ of the vector $\cal X$ are given in the following form
\begin{equation}
\label{diagonalx}
X_{i}= \left( \frac{2}{i(i+1)}\right)^{{1}/{2}}\left( (i+1)P_{ii}+\sum_{j=i+1}^{N-1}P_{jj}-r \right), \quad i=1,\cdots, N-1,
\end{equation}
where we see that $X_{1}$ could also be written as $X_{1}= P_{11}-P_{NN}$.
Conversely, we can easily see that we have
\begin{equation}
{\cal P}= A^{-1} {\cal X} + \frac{r}{N} \left(\begin{array}{c}
1 \\
\vdots \\
1
\end{array}\right),
\end{equation}
so that the diagonal entries of $P$ can be expressed as the following linear combination of the $X_1,\cdots, X_{N-1}$:
\begin{eqnarray}
&&P_{ii}=\left( \frac{i}{2(i+1)}\right)^{{1}/{2}}X_{i}-\frac{1}{2} \sum_{j=i+1}^{N-2} \left( \frac{2}{j(j+1)}\right)^{{1}/{2}}X_{j}+\frac{r}{N} ,  \nonumber \\
&&P_{NN}= r- \sum_{j=1}^{N-1}P_{jj}, \quad i=1,\cdots, N-1, .
\label{diagonalp}
\end{eqnarray}

The relative freedom in the choice of the components of the vector ${\cal X}$ is reflected in (\ref{inreachingcanexp4}). Indeed, if we take ${\cal X'}=S {\cal X}= A' {\cal P}+b' $, our canonical choice will be preserved if $S$ is a real orthogonal matrix. The constant $C(N,r)$ thus remains invariant.
If $S$ is not orthogonal, we see that both the equation of the surface and the constant $C(N,r)$ are different (non canonical).


\section{Independent coordinates of the surface in the $\field{C}P^{N-1}$ model
\label{coordsurfacesrk1}}

In the previous section, we have shown how the coordinates of the vector $X \in \field{R}^{N^2-1}$ could be  obtained from an arbitrary projector $P$ constructed from solutions of the $\field{C}P^{N-1}$ sigma model. We have also given a canonical quadratic equation (\ref{forsrfaceeqfinalform})
of a surface in $\field{R}^{N^2-1}$ satisfied by these coordinates. 

Let us mention that (\ref{forsrfaceeqfinalform}) has been obtained by adding the $N$ equations (\ref{genconst1}) so more constraints are imposed on the  coordinates of the vector $X$ from the condition $P^2=P$ and the question which arises now is how to find the independent coordinates and what  are they? To our knowledge, the answer is known only in the cases of rank 1 and rank $(N-1)$. Indeed, both cases give rise to $2(N-1)$ real independent quantities. 

Indeed if we take $P$ as an orthogonal projector of rank 1, it could be written as 
\begin{equation}
P= {\hat u}\otimes {\hat u}^\dagger, \quad  {\hat u}^\dagger \cdot {\hat u}=1, \quad u \in \field{C}^{N}.
\label{rank1projector}
\end{equation}
 It is thus characterised by $(2N-1)$ real independent quantities which could be chosen as the entries $P_{1i}$, where $ i=1,...,N$, and 
\begin{equation}
{\hat u}^T=\frac{1}{\sqrt{P_{11}}}\ (P_{11},\ P_{12},..., P_{1N}).
\label{vectoru}
\end{equation}
The other entries of $P$ are given by 
\begin{eqnarray}
\begin{array}{c}
P_{ii}=\frac{|P_{1i}|^2}{P_{11}}\,, \\
\, \\
P_{ij} = \frac{{\bar P}_{1i}P_{1j}}{P_{11}}\,,
\end{array}
\qquad
\begin{array}{c}
i = 2, \ldots, N, \\
j = 3, \ldots, N,
\end{array}
\qquad
i<j\,.
\label{constfork2}
\end{eqnarray}
Using (\ref{xijplusminus}) and (\ref{diagonalx}) , we can write the explicit constraints on the coordinates of the radius vector $X$ of  $\field{R}^{N^2-1}$, which enable us to show that the $(N-1)$ coordinates $X_i$ and the $(N-1)(N-2)$ coordinates $\{(X_{ij})_+, \ (X_{ij})_-, \ i<j,  \ i=2, \ldots, N, \ j=3, \ldots, N\}$
are related to the $(2N-1)$ real independent coordinates $\{P_{11}, \ (X_{1j})_+, \ (X_{1j})_-, \ j=2, \ldots, N\}$. Indeed, we get the following expressions:
\begin{eqnarray}
X_1&=& (2 P_{11}-1)+ \frac{1}{4 P_{11}}{\sum_{j=i+1}^{N-1}\left( (X_{1j})_+^2+(X_{1j})_-^2\right)} ,\nonumber\\
X_{i} &=&  \frac{1}{4 P_{11}} \left( \frac{2}{i(i+1)}\right)^{{1}/{2}}
\Bigg( (i+1) \left( (X_{1i})_+^2+(X_{1i})_-^2\right)  \nonumber \\
&+&\sum_{j=i+1}^{N-1}\left( (X_{1j})_+^2+(X_{1j})_-^2\right)- 4 P_{11} \Bigg), 
\end{eqnarray}
for $i=2,\ldots, N-1$ and 
\begin{eqnarray}
 (X_{kl})_+ &=& \frac{1}{2 P_{11}} \left(  (X_{1k})_+  (X_{1l})_+ + (X_{1k})_-  (X_{1l})_-\right),   \nonumber \\
  (X_{kl})_- &=& \frac{1}{2 P_{11}} \left(  (X_{1k})_+  (X_{1l})_- - (X_{1k})_-  (X_{1l})_+\right),
\label{constraintdiagonalx}
\end{eqnarray}
for $k<l,\ k=2,\ldots, N,\ l=3,\ldots, N.$

Let us mention that the case of a projector of rank $(N-1)$, say $Q$, is easily deduced from the case of rank one. Indeed, due to the completeness relation (\ref{completenessrel}), we get $Q=I-P$, where $P$ is an orthogonal projector of rank one. We thus find from (\ref{constfork2}), similar expressions on the entries of $Q$:
\begin{eqnarray}
\begin{array}{c}
Q_{ii} = 1+\frac{|Q_{1i}|^2}{Q_{11}-1}\,, \\
\, \\
Q_{ij} = \frac{\bar{Q}_{1i}Q_{1j}}{Q_{11}-1}\,,
\end{array}
\qquad
\begin{array}{c}
i = 2, \ldots, N, \\
j = 3, \ldots, N,
\end{array}
\qquad
i<j\,.
\label{indconstfornminus1}
\end{eqnarray}


\section{Special solutions of the $\field{C}P^{N-1}$ model and projectors
\label{verseq}}

We will now consider special solutions of the $\field{C}P^{N-1}$ model obtained from the holomorphic vectors corresponding to the Veronese sequence. It is well-known that the Veronese vector $f$ for the $\field{C}P^{N-1}$ sigma model 
can be written as
\begin{eqnarray}
f&= &(w_0^0, w_1^0, \dots, w_{N-1}^0)^T\nonumber\\
&=&\left(1, \sqrt{\left(\begin{array}{c}
N-1 \\
1 \end{array}\right)}\, \xi\,, \dots, 
\sqrt{\left(\begin{array}{c}
N-1 \\
r \end{array}\right)}\, \xi^r\,, \dots,
\xi^{N-1}
\right)^T\, .
\label{veroneseseq}
\end{eqnarray}

Following the procedure described in Section \ref{briefintroclass}, we construct the mixed solutions 
given by $P_{+}^k f$. Explicitly, we get $N-2$ mixed solutions that are given as
\begin{equation}
P_{+}^k f= (w_0^k, w_1^k, \dots, w_{N-1}^k)^T, \quad k= 1, \ldots, N-2,
\end{equation}
where
\begin{eqnarray}
w_r^k &=& \frac{1}{(1+|\xi|^2)^k}  \sqrt{\left(\begin{array}{c}
N-1 \\
r \end{array}\right)}\,\xi^{r-k}\alpha^k_r,
\label{mostgenmixedsol}
\end{eqnarray}
and where, in turn, 
\begin{equation}
\alpha^k_r\,=\,\frac{1}{(r+1)} \frac{1}{(r-N)}\sum_{l=0}^k  \vert \xi\vert^{2l}
\left(\begin{array}{c}
k \\
l \end{array}\right)\prod_{i=0}^{l} (r+i-N) \prod_{j=l}^k(r-k+j+1).
\label{mostgenmixedsol1}
\end{equation}

We can thus easily get the following expressions,  $k=0,...,N-2$ 
 \begin{equation}
{\vert P_{+}^k f \vert^2}=\frac{(N-1)! \ k!}{(N-(k+1))!} \left(1+|\xi|^2\right)^{N-1-2k}\,,
\label{normfork}
\end{equation}
\begin{equation}
(P_{+}^k f)^{\dagger} \cdot \  \partial P_{+}^k f = \frac{(N-1)! \ k!}{(N-(k+1))!} \big(N-(2k+1)\big) \bar{\xi} \left(1+|\xi|^2\right)^{N-1-(2k+1)}\,
\label{normforpartialk}\,.
\end{equation}

Due to our way of constructing  mixed solutions, we can provide a formula (in the special case of the Veronese sequence) which relates two such solutions, say,  $P_{+}^{(N-1-k)}f$ 
and $P_{+}^k f$ . Indeed, we have,
\begin{eqnarray}
P_{+}^{(N-1-k)}f = (-1)^k \gamma_k^N(|\xi|^2) \mathcal{A} \overline{P_{+}^k f}\,, \qquad k=1,\ldots, N-2\,,
\label{relationbetfs}
\end{eqnarray}
where $\gamma_k$ is the ratio of the normalization factors 
\begin{eqnarray}
\gamma_k^N(|\xi|^2) &=& \frac{{\vert P_{+}^{NÊ-1-k} f \vert^2}}{{\vert P_{+}^k f \vert^2}} \nonumber \\
&=& \frac{(N-1-k)!}{k!} \left(1+|\xi|^2\right)^{-N+1+2k}\,
\label{rationorfac}
\end{eqnarray}
and $\mathcal{A}$ is an $N \times N$ anti-diagonal matrix whose non-zero elements are given by 

\begin{eqnarray}
\mathcal{A}_{j(N-j+1)} = (-1)^{N+j}\,, \qquad j=1,\ldots, N\,.
\label{matrixforrelN}
\end{eqnarray}

We could thus conclude that the following Veronese vectors are related (in the same sense as before,
{\it ie} the two vectors are related if one of them is the complex conjugate of the other,
and their entries are $\pm$ of each other taken in a reverse order)
\begin{eqnarray}
&&P_+^{N-2}f \propto P_+f\,, \nonumber \\
&&P_+^{N-3}f \propto P_+^2f\,, \nonumber \\
&&\vdots \nonumber \\
&&\left\{
\begin{array}{c}
P_+^{\frac{N-2}{2}+1}f \propto P_+^{\frac{N-2}{2}}f\,, \quad {\rm if}\,\,\, N-1\,\,\,\,{\rm is\,\, odd}\,, \\
P_+^{\frac{N-1}{2}}f \propto P_+^{\frac{N-1}{2}}f\,, \qquad\, {\rm if}\,\,\, N-1\,\,\,\,{\rm is\,\, even}\,,
\end{array}\right.
\label{finalrelations}
\end{eqnarray}
In particular for even values of $N-1$ we have 
\begin{eqnarray}
P_+^{\frac{N-1}{2}}f = (-1)^{\frac{N-1}{2}} \mathcal{A}\ \overline{{P_+}^{\frac{N-1}{2}}f}\,,
\label{partirelforeven}
\end{eqnarray}
where $N \times N$ anti-diagonal matrix $\mathcal{A}$ becomes  
\begin{eqnarray}
\mathcal{A}_{j(N-j+1)} = (-1)^{1+j}\,, \qquad j=1,\ldots, N\,.
\label{matrixforrelNeven}
\end{eqnarray}


\section{Reduced projectors}

Hence, for even values of $(N-1)$ we have the so called "reduced"
projector $P_{\frac{N-1}{2}}$ constructed out of Veronese 
vectors $P_+^{\frac{N-1}{2}}f$, which is symmetric with respect to the 
anti-diagonal elements with proper $\pm$ signs. Naturally this kind 
of a projector has more constraints between its entries. Let us now concentrate on the 
off-diagonal elements. The number of the additional constraints 
can be given as
\begin{eqnarray}
&&N-2\,, \qquad {\rm coming\,\,\, from\,\,\, the\,\,\, first\,\,\, row}\,, \nonumber \\
&&N-4\,, \qquad {\rm coming\,\,\, from\,\,\, the\,\,\, second\,\,\, row}\,, \nonumber \\
&&N-6\,, \qquad {\rm coming\,\,\, from\,\,\, the\,\,\, third\,\,\, row}\,, \nonumber \\
&&\vdots \nonumber \\
&&N-(N-1)\,, \qquad {\rm coming\,\,\, from\,\,\, the\,\,\,} \frac{N-1}{2} {\rm th\,\,\, row}\,. 
\label{additionalconst}
\end{eqnarray}
Previously, by applying a standard procedure and considering only the 
off-diagonal elements we have obtained $(N^2-N)$ components for the radius vector 
(i.e. $(X_{ij})_\pm$ as in (\ref{xijplusminus})). In order to find how many components has the 
radius vector corresponding to this reduced projector we need to substract 
two times the sum of the additional constraints given in 
(\ref{additionalconst}) from $(N^2-N)$ which will give us $\frac{N^2-1}{2}$. 

It is clear that the number of the components of the radius 
vector obtained from the diagonal elements of the reduced 
projector is $\frac{N-1}{2}$, since the reduced projector 
is symmetric with respect to the anti-diagonal elements. However, 
let us justify this for the canonical choice of the components of the 
radius vector on examples. We will consider the $\field{C}P^{4}$ and 
$\field{C}P^{6}$ cases ($\field{C}P^{2}$ case is trivial, since for 
$P_1$ we only have $X_2$).

{\bf i) $\field{C}P^{4}$ case:}

For this case the reduced projector is $P_2$
and the linear combination of $X_i$'s, obtained from this projector 
($X_1=P_{11}-P_{NN}=0$) 
\begin{equation}
\sum_{i=2}^4 a_iX_i = 0\,,
\label{linsumcp4}
\end{equation}
is satisfied for 
\begin{equation}
a_2=-2\sqrt{\frac{2}{15}}\, a_4\,, \qquad a_3 = \frac{a_4}{\sqrt{15}}\,. 
\end{equation}
Hence, $X_2$ could be expressed in terms of $X_3$ and $X_4$
\begin{equation}
X_2 = \frac{1}{2\sqrt{2}}X_3 + \frac{\sqrt{15}}{2\sqrt{2}}X_4\,, 
\end{equation}
which justifies our claim that we have only two independent diagonal components.

{\bf ii) $\field{C}P^{6}$ case:}

For this case the reduced projector is $P_3$
and the linear combination of $X_i$'s, obtained from this projector  
\begin{equation}
\sum_{i=2}^6 a_iX_i = 0\,,
\label{linsumcp6}
\end{equation}
is satisfied for 
\begin{equation}
a_2=-\frac{2a_6}{\sqrt{7}}\,, \qquad a_3 = -\frac{1}{2}\sqrt{\frac{5}{2}}\,a_5 + \frac{3a_6}{2\sqrt{14}}\,, 
\qquad a_4 =\frac{a_5}{2\sqrt{6}} + \frac{1}{2}\sqrt{\frac{5}{42}}\,a_6\,.
\end{equation}
Hence, $X_2$ and $X_3$ could be expressed in terms of $X_4$, $X_5$ and $X_6$,
\begin{eqnarray}
&&X_2 = \frac{2}{\sqrt{15}}X_4 + \frac{3}{2\sqrt{5}}X_5 + \frac{\sqrt{7}}{2} X_6\,, \nonumber \\
&&X_3 = \frac{1}{\sqrt{15}}X_4 + \frac{2\sqrt{2}}{\sqrt{5}}X_5\,,
\end{eqnarray}
which shows that we have three independent diagonal components. 

\subsection{$\field{C}P^{2n-1}$ case}

This is the case when $N=2n$ is even. As a consequence of ({\ref{relationbetfs}),  let us note that the vector $f$ is related to $P_+^{2n-1}f$
and in general $P_+^{k}f$ to $P_+^{2n-k-1}f$. This has interesting implications
for the projectors $P_k$ and in particular, for some of their sums.
Thus, in particular the expressions are very simple for sums of two projectors
$P_k+P_{2n-k-1}$ for $k=0,...,n-1$.
In each case the resultant projector has the same structure. Assuming that it is
given by a $2n\times 2n$ matrix $C$, all  its diagonal 
entries satisfy $C_{i,i}=C_{2n-i,2n-i}$.
As the projector is hermitian, its off-diagonal entries satisfy
$C_{i,j}=\bar C_{j,i}$. Furthermore, all elements $C_{i,2n-i+1}$ vanish.
Moreover, there are various symmetries amongst the entries along the lines parallel
to the diagonal. Indeed, we have  $C_{i,i+1}=-C_{2n-i,2n-i+1}$, $C_{i,i+2}=C_{2n-i-1,2n-i+1}$ 
and in general $C_{j,j+k}=(-1)^{k} C_{2n-k-j+1,2n-j+1},\quad j=1,...,2n$. Thus all entries in matrix $C$
are determined by the independent entries shown below

\begin{eqnarray}
 \left(\begin{array}{cccccccccc}
A_{1,1}&a_{1,2}&a_{1,3}&\cdots&a_{1,n}&a_{1,n+1}&\cdots&a_{1,2n-2}&a_{1,2n-1}&0 \\
\,&A_{2,2}&a_{2,3}&\cdots&a_{2,n}&a_{2,n+1}&\cdots&a_{2,2n-2}&0&\, \\
\,&\,&\ddots&\,&\vdots&\vdots&\,&\,\iddots&\,&\,\\
\,&\,&\,&\,&a_{n-1,n}&a_{n-1,n-1}&\,&\,&\,&\,\\
\,&\,&\,&\,&A_{n,n}&0&\,&\,&\,&\,\\
\end{array}\right)\,,
\label{part}
\end{eqnarray}
where $A_{i,i}$ are real and $a_{i,j}$ are complex.

All the other entries are determined in terms of these.
Hence the total number of independent quantities is $n$ (real) + $n(n-1)$ (complex) - 1 (due to the trace)
= $(n-1)(2n+1)$.

The same is true if we take sums of the pairs of projectors $P_i+P_{2n-i-1}$,{\it ie}
$P_1+P_{2n-2}$ +, say, $P_3+P_{2n-4}$. We can also take $\sum_{i=1}^{2n-1} P_i$.

\subsection{More on the $\field{C}P^{2n-2}$ case}

This is the case when $N=2n-1$ is odd. We can perform a similar discussion and each vector $P_+^kf$ has $(2n-1)$ components.The projector $P_{n-1}$, as mentioned at the begining of this Section, is very special, as it by itself, describes
the "reduced" case. The symmetries mentioned before show that its diagonal terms
satisfy $P_{i,i}=P_{2n-1,2n-1}$ and, of course, the entry $P_{n,n}$ has no `partner' and so
stands by itself. The off-diagonal terms satisfy relations similar to what we had in the case when $N-1$ is odd, except that, for each line parallel to the diagonal, there is an unpaired entry.
Hence, like in the odd case all the entries of the matrix representing  $P_{n-1}$ are determined in terms of a matrix whose independent entries are given by 
 
\begin{eqnarray}
 \left(\begin{array}{ccccccccc}
A_{1,1}&a_{1,2}&a_{1,3}&\cdots&a_{1,n}&a_{1,n+1}&\cdots&a_{1,2n-2}&a_{1,2n-1} \\
\,&A_{2,2}&a_{2,3}&\cdots&a_{2,n}&a_{2,n+1}&\cdots&a_{2,2n-2}&\, \\
\,&\,&\ddots&\,&\vdots&\vdots&\,&\iddots\,&\,\\
\,&\,&\,&\,&A_{n,n}&a_{n,n+1}\,&\,&\,&\,\\
\end{array}\right)\,,
\label{parta}
\end{eqnarray}
where, again, $A_{i,i}$ are real and $a_{i,j}$ are complex.

As before, the same can be said about the structure of the pairs of projectors
 $P_i$ and $P_{2n-i}$. Moreover, we can also take sums of such pairs of projectors
 and add to it the projector $P_{n-1}$.
 The number of independent entries, in this case, is the same as in the odd case mentioned 
 before.
 
 \subsection{Examples}
 
Let us demonstrate these claims on a specific example when $n=2$.
The projector $P_1$ for the $\field{C}P^2$ model is proportional to
\begin{equation}
\label{NEW1}
\left(\begin{array}{ccc}
4\vert \xi\vert^2& -2\sqrt{2}\bar \xi(1-\vert \xi\vert^2)& 4\bar \xi^2\\
-2\sqrt{2} \xi(1-\vert \xi\vert^2)&2(1-\vert \xi\vert^2)^2&2\sqrt{2}\bar \xi(1-\vert \xi\vert^2)\\
-4\xi^2&2\sqrt{2}\xi(1-\vert \xi\vert^2)&4\vert \xi\vert^2\\
\end{array}\right),
\end{equation}
while the sum of projectors $P_1+P_2$ of the $\field{C}P^3$ model ({\it i.e.} a solution of the G(4,2) model) is proportional to
\begin{equation}
\label{NEW2}
\left(\begin{array}{cccc}
3\vert \xi\vert^2(1+\vert \xi\vert^2) & -\sqrt{3}\bar \xi(1-\vert \xi\vert^4)& -\sqrt{3}\bar \xi^2(1+\vert \xi\vert^2)&0\\
-\sqrt{3} \xi(1-\vert \xi\vert^4)&1+\vert \xi\vert^6&0& -\sqrt{3}\bar \xi^2(1+\vert \xi\vert^2)\\
-\sqrt{3} \xi^2(1+\vert \xi\vert^2)&0&1+\vert \xi\vert^6& \sqrt{3}\bar \xi(1-\vert \xi\vert^4)\\
0&-\sqrt{3} \xi^2(1+\vert \xi\vert^2)&\sqrt{3}\xi(1-\vert \xi\vert^4)&3\vert \xi\vert^2(1+\vert \xi\vert^2)\\
\end{array}\right).
\end{equation}
Clearly, both expressions lead to surfaces in $\mathbb{R}^5$ as in each case we have 2 independent complex entries
(say, second and third entries of the first rows) and one from the diagonals (if we impose the condition of the trace).
Similar expressions can be given for higher dimensional cases.

 Thus we see that the reduced projectors 
 of the  $\field{C}P^{2n-2}$ and $\field{C}P^{2n-1}$ models have the same number
 of independent entries and so, lead to vectors $X$ in the space of the same 
 number of dimensions.

\subsection{Properties of these solutions}

As we have found that, for maximally reduced cases, the number of independent variables
of surfaces based on some projectors in $CP^{N-1}$ models (odd and even cases of $N$)
are the same. The question then arises as to whether these surfaces are really the same.

One way to study this involves the consideration of the curvatures of these surfaces.
To do this we need to calculate the metric on these surfaces
\begin{equation}
\label{metric}
g_{++}\,=\,\hbox{tr}
(\partial P\partial P),\quad g_{+-}\,=\,\hbox{tr}
(\partial P\bar\partial P),\quad g_{--}
\,=\,\overline{g_{++}},
\end{equation}
where $P$ is the corresponding projector (or a sum of projectors) 
and then calculate the Gaussian curvature $K$.

However, due to the orthogonality of $P_+^kf$ vectors, we have $g_{++}=g_{--}=0$
and the only nonvanishing component of the metric is $g_{+-}$.

The Gaussian curvature $K$ is then given by (\ref{curve}).

Taking the now a general sum of  projectors $P_i$ as given in (\ref{concproject}), we get $P=\sum_{i=0}^{k} \alpha_i P_i$ where $\alpha_i\in \mathbb R$. It is thus easy  to check using (\ref{trace1}) that $g_{+-}$  is given by 
\begin{equation}
\label{sums}
g_{+-}=\alpha_0^2 \frac{\vert P_+ f\vert^2}{\vert f \vert^2}+\sum_{i=1}^{k} \alpha_i^2 \big(\frac{\vert
P_+^{i+1}f\vert^2}{\vert P_+^{i}f
\vert^2}+\frac{\vert
P_+^i f\vert^2}{\vert P_+^{i-1}f
\vert^2}\big)- 2 \sum_{i=0}^{k-1} \alpha_i \alpha_{i+1} \frac{\vert
P_+^{i+1}f\vert^2}{\vert P_+^{i}f \vert^2}.
\end{equation}

However, for the Veronese sequence, we can show from (\ref{normfork}) that such ratios are  given by
\begin{equation}
\label{prop}
\frac{\vert
P_+^if\vert^2}{\vert P_+^{i-1}f
\vert^2}\,=\,\frac{i(N-i)}{(1+\vert \xi\vert^2)^2},
\end{equation}
 so  the final expression for the metric is 
 \begin{equation}
\label{finmetric}
g_{+-}=\frac{A(N,P)}{(1+\vert \xi\vert^2)^2},
\end{equation}
where the constant $A(N,P)$ depends on $N$ but also on $P$ (in fact, on the $\alpha_i's$). It is explicitly given by 
 \begin{equation}
A(N,P)=\sum_{i=0}^{k-1}(i+1)(N-(i+1))(\alpha_i-\alpha_{i+1})^2 + (k+1)(N-(k+1)) \alpha_k^2.
\end{equation}

Since, we have
\begin{equation}
\label{cur}
\partial \bar\partial \ln\left(\frac{A}{(1+\vert \xi\vert^2)^2}\right)\,=\,-2\frac{1}{(1+\vert \xi\vert^2)^2},
\end{equation}
for any constant $A$, we see that the Gaussian curvature is proportional to
\begin{equation}\label{curva}
K\,=\,\frac{8}{A(N,P)},
\end{equation}
{\it ie} is constant. Hence all our surfaces have a constant curvature. So are they the same?

To check this we have looked at the two simplest cases, namely, of the projector $P_1$ of $CP^2$ model
and of the sum of the projectors $P_1+P_2$ of $CP^3$, {\it i.e.} the example discussed in the previous subsection.
The first case give us $A(3,P_1)=4$ while the second gave us $A(4, P_1+P_2)=6$ and we conclude
that the curvatures are different. So these surfaces are really in the same dimensional spaces
and both have constant curvatures, but their curvatures are different.

\section{Conclusions and final comments}

In this paper we have looked at various properties of projectors representing harmonic maps of $S^2$ into grassmannians.
Some of these properties are very well known (see {\it e.g.} Ref.9 and references therein) or are probably known to 
many people but we have assembled them here as we can use these projectors for the generation of various
surfaces in multidimensional spaces based on these maps. In particular, we looked in detail on the projectors
corresponding to the Veronese sequence of such maps.

For the Veronese sequence of maps some projectors have many symmetries and some have vanishing entries. These symmetries lead
to the generated surfaces lying in lower dimensional spaces. The projectors corresponding to the lowest such spaces were called "reduced" in this paper. We also showed that there are some simple relations between such projectors for $CP^{N-1}$ harmonic  
maps corresponding to $N$ being odd and even. 

We have also discussed in detail the construction of surfaces. In particular we have showed that there are at least two ways in constructing such surfaces; one involving line integrals of derivatives of projectors (this approach was used in previous
studies)  and on the direct use of the projectors. This last (and new) approach generates more surfaces
and agrees with the first one when the harmonic maps are selfdual (see (\ref{newc}) and (\ref{newd})).

We have also showed that, in general, the surfaces have nonconstant Gaussian curvature of the induced metric but that 
for the Veronese sequence all metrics are proportional to each other and the curvature is constant.
However, the value of this constant still depends on the map itself.

\section*{Acknowledgments}
This work is supported in part by supported by research grants  from NSERC of Canada. IY acknowledges a postdoctoral fellowship awarded by the laboratory of mathematical physics of the CRM, Universit\'e de Montr\'eal. This paper was finished when VH visited the University of Durham during the academic year 2009-2010.


\end{document}